\documentstyle[prd,aps,epsfig,preprint,tighten]{revtex}

\begin{document}
\draft
\title{		PARAMETRIZATION OF THE ENERGY SPECTRUM IN THE TRITIUM BETA DECAY.}
\author{	J. STUDNIK AND M. ZRA\L EK}
\address{      Department of Field Theory and Particle Physics, 
\\
                Institute of Physics, University of Silesia, 
\\
                Uniwersytecka 4, PL-40-007 Katowice, Poland  	}
\maketitle

\begin{abstract}

Taking into account  
mixings among neutrino states, the end of the energy spectrum of the electron in the tritium beta decay is investigated. 
It is shown that for real energy resolutions of a spectrometer, $ \Delta E > 0.08$ eV, the effective electron neutrino mass 
should [not] be taken in the  
form
\begin{center}
 $m_{\beta}^{(1)}=\sqrt{\sum|U_{ei}|^2 m_i^2}$  
$ \ \ \ \ \ \ \left[ m_{\beta}^{(2)}=\sum|U_{ei}|^2 m_i \right]$.
\end{center}

\vspace{1cm}

\end{abstract}

Last atmospheric and solar experiments convince us that neutrinos are massive particles. However, the problem of absolute values of 
their masses is still waiting for a solution. Apparently  three kinds of neutrino experiments have a chance to determine the light neutrino masses.
These are:
\begin{enumerate}
\item neutrino oscillation experiments, 
\item the tritium beta decay experiment, 
\item neutrinoless double beta decay experiments.
\end{enumerate}

Among them the tritium $\beta$ decay
\begin{equation}
^3_1H \rightarrow ^3_2He+e^-+\bar{\nu _e},
\end{equation}
is of great importance. Here the end of the electron energy spectrum in the absence of the lepton mixing is described by \cite{neutphys} 

\begin{equation}
\label{delta}
\frac{dN}{dE}=R(E)(E_0-E) \sqrt{(E_0-E)^2-m_{\beta}^2}.
\end{equation}

\hspace{-6mm}
$E$ is the electron kinematic energy $(E=E_{tot}-m_{e} \approx \frac{p^2}{2m_{e}})$, $E_0=M(^3_1 H)-M(^3_2 He)-m_e \approx 18572.1$ eV,
and

\begin{equation}
R(E)=G_F^2 \frac{m_e^5 cos \theta_c}{2 \pi ^3} |M|^2 F(E) \sqrt{2m_e E}(E+m_e),
\end{equation}
where $G_F$ is the Fermi constant, $\theta_c$ is the Cabibbo angle and M is the nuclear matrix element. $F(E)$ is  neutrino mass independent, 
smooth function of $E$, which takes into account radiative corrections to the produced final state electron 
(see \cite{neutphys} for details). The effective mass $m_{\beta}$ of the produced neutrino is determined, 
regardless of the fact if it  is a Dirac or a Majorana particle. 

Presently only the upper limit on $m_{\beta}$ is available. Two experiments in Mainz \cite{mainz} and Troitsk \cite{troisk} have  recently found
\begin{eqnarray}
\label{mt}
m_{\beta} &<& 2.2  \ eV \cite{mainz}, \\ \nonumber
m_{\beta} &<& 2.5 \  eV \cite{troisk}.
\end{eqnarray}

The real problem is how this effective mass $m_{\beta}$, extracted from the experiment, is connected to the realistic neutrino masses $m_i$. 
In the case of three light neutrinos (i=1,2,3) which mix ($\left| \nu_e \right> = \sum U_{ei} \left| \nu_i \right>$), 
the spectrum of the electron energy is given by \cite{eg} 

\begin{equation}
\label{delta1}
\frac{dN}{dE}=R(E)(E_0-E)\sum_{i=1}^3|U_{ei}|^2 \sqrt{(E_0-E)^2-m_{i}^2}  \ \Theta (E_0-E-m_{i}).
\end{equation}

$ \Theta (E_0-E-m_{i})$ is the step function.
This formula 
depends on five parameters, namely three neutrino masses and two mixings because $\sum |U_{ei}|^2=1$.
 If only the upper bound on $m_{\beta}$ is determined (present situation, Eq.~\ref{mt}) then a precise relation
\begin{equation}
\label{fun}
m_{\beta}=f(|U_{ei}|^2,m_i)
\end{equation}
is not so important. However, future experiments, as KATRIN   \cite{katrin} with the sensitivity $m_{\beta} \sim 0.3 - 0.35 \ eV$
have a chance to find $m_{\beta} \neq 0$ and then a form of Eq.~\ref{fun} becomes 
crucial for neutrinos mass determination. 
 Two parametrizations have been  suggested in literature. 
The first one \cite{vissani}
\begin{equation}
\label{vis}
m^{(1)}_{\beta}=\sqrt{\sum_{i=1}^3|U_{ei}|^2 m_i^2}.
\end{equation}
is a result of Taylor expansion of the spectrum (~\ref{delta1}) around the point 
\begin{equation}
\left( \frac{m_i}{E_0-E} \right)^2 \approx 0.
\end{equation}
The second \cite{farzan}
\begin{equation}
\label{far}
m^{(2)}_{\beta}=\sum_{i=1}^3|U_{ei}|^2 m_i,
\end{equation} 
follows from the approximation of the precise distribution (Eq.~\ref{delta1}) by the effective one (Eq.~\ref{delta}) near
the end of the electron energy spectrum $E \rightarrow E_0$. We would like to elucidate the situation  and decide which parametrization, 
(~\ref{vis}) or (~\ref{far}) is better and should be used for future neutrinos mass determination.

Among five parameters $|U_{ei}|$ and $m_i$ only one is actually unknown, it is the mass of the lightest one $m_1$. 
From neutrino oscillation experiments the mixing matrix elements $|U_{ei}|^2$, (i=1,2,3) $\delta m^2_{atm}=m_3^2-m_2^2$ and $\delta m^2_{solar}=m_2^2-m_1^2$ 
are determined. For the LMA MSW solution of the solar neutrino problem the best fit values are \cite{gonzales}
\begin{equation}
|U_{e1}|^2=0.55, \ |U_{e2}|^2=0.43, \ |U_{e3}|^2=0.02,
\end{equation}
and $\delta m^2_{solar}=3.5 \times 10^{-5} \ eV^2$. Atmospheric neutrino oscillations give $\delta m^2_{atm}=3.1 \times 10^{-3} \ eV^2$ \cite{gonzales}, 
\cite{toshito}. Uncertainties in the determination of the oscillation parameters from solar neutrino experiments are large \cite{gonzales}, 
but our main conclusions depend only weakly on them. The masses of heavier neutrinos are the function of the lightest neutrino $m_1$
\begin{eqnarray}
m_2&=&\sqrt{m_1^2+\delta m^2_{solar}}, \\ \nonumber
m_3&=&\sqrt{m_1^2+\delta m^2_{solar}+\delta m_{atm}^2}.
\end{eqnarray}
As $R(E)$ is a smooth function of $E$ at the end of $\beta$ spectrum, 
we can approximate $R(E) \approx R(E_0-m_1)$. Then we can plot scaled energy distribution as:
\begin{eqnarray}
\frac{1}{R(E_0-m_1)} \frac{dN}{dE} \equiv f_i(E).
\end{eqnarray}

\begin{figure}[h]
\begin{center}
\epsfig{figure=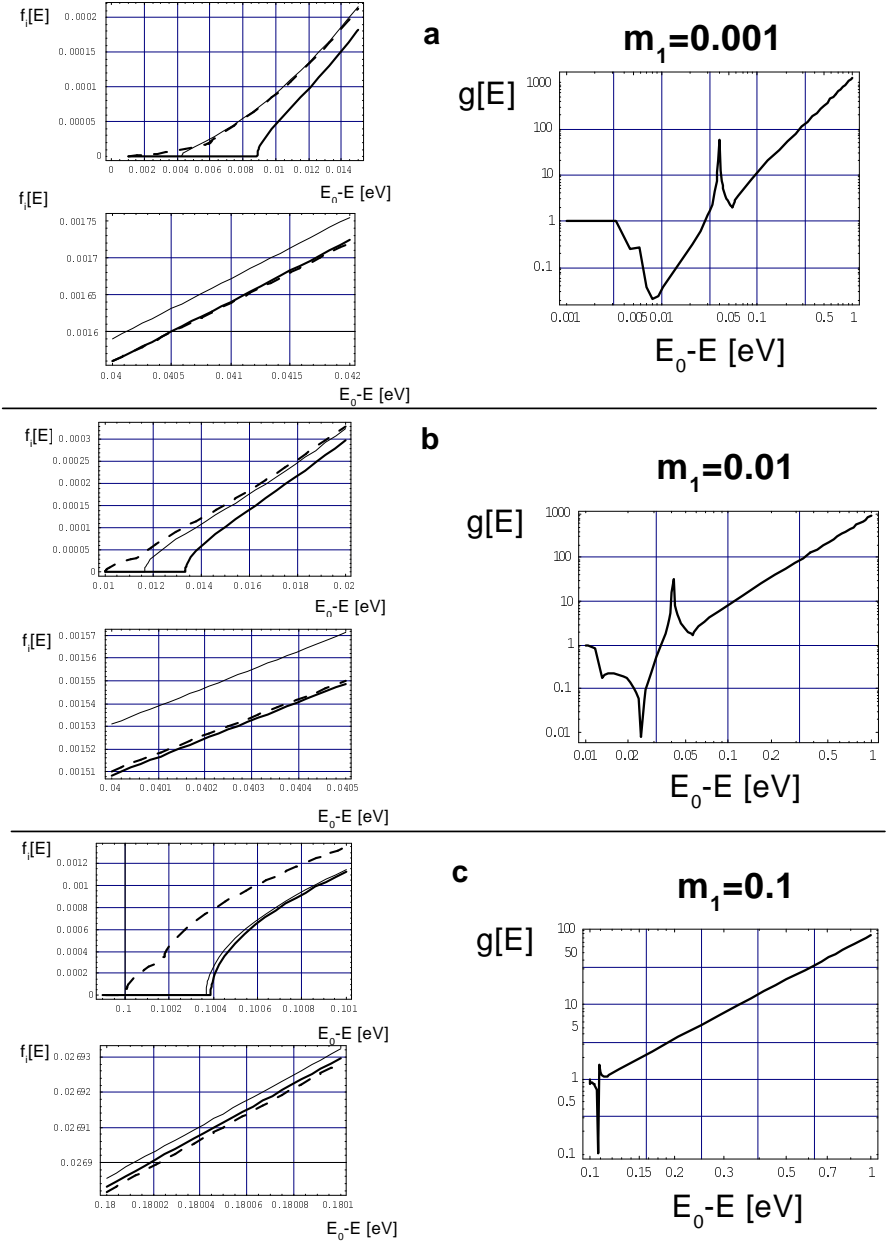, scale=0.6}

\caption{The scaled electron energy distribution $f_i(E)$ at the end of the spectrum for $^3_1H$ decay for three different masses of the lightest neutrino (a) $m_1=0.001$ eV, (b) $m_1=0.01$ eV, (c) $m_1=0.1$ eV. $f_0(E)$ descibes the full (dashed line) and $f_i(E)$ (i=1,2) describes approximate effective energy distribution for $m_{\beta}^{(1)}$ (tick solid line) and  $m_{\beta}^{(2)}$ (thin solid line). The function g(E) compare both approximations (see text).}
\label{g}
\end{center}
\end{figure}

In Fig.~\ref{g} the full $f_0(E)$ (Eq.~\ref{delta1}) and two effective distributions 
( $f_1(E)$ with $m_{\beta}=m_{\beta}^{(1)}$ and  $f_2(E)$ with $m_{\beta}=m_{\beta}^{(2)}$)  are depicted as a function  of energy $E_0-E$ for three particular values of the lightest neutrino 
masses ($m_1=0.001 \ eV, m_1=0.01 \ eV $ and $ m_1=0.1 \ eV$). 
To compare both approximations, the ratio
\begin{equation}
g(E)=\frac{|f_0(E)-f_2(E)|}{|f_0(E)-f_1(E)|}
\end{equation}
is also shown.
We can see that for small values of $x=E_0-E$ the effective distribution $f_1(E)$ with $m_{\beta}^{(1)}$ approximates the full spectrum in a better way
$(g(x) <1)$. For larger $x$, $g(x)>1$, and  $m_{\beta}^{(2)}$ gives better result. This conclusion is general, independent of the lightest neutrino mass $m_1$ 
and values of the other oscillation parameters. 
To answer the question which effective neutrino mass $m_{\beta}^{(1)}$ or $m_{\beta}^{(2)}$ should be used in future experimental searches,
 the number of events in a possible small  interval $\Delta E$ which still can be resolved by a detector

\begin{equation}
\left(E_0-m_1- \Delta E , E_0-m_1 \right)
\end{equation}
should be calculated. The integral
\begin{equation}
n_i(\Delta E)=\int_{E_0-m_1- \Delta E}^{E_0-m_1} f_i(E) \delta E
\end{equation}
can be done analytically,

\begin{eqnarray}
n_0(\Delta E)= \frac{1}{3R(E_0-m_1)}\{ |U_{e1}|^{2}B^{3/2}+|U_{e2}|^2 \left( B- \delta m^2_{solar} \right)^{3/2} \times \\ \nonumber
\times \Theta \left (\Delta E-(m_2-m_1) \right)+|U_{e3}|^2 \left( B- \delta m^2_{solar}-\delta m_{atm}^2 \right)^{3/2} \Theta \left (\Delta E-(m_3-m_1) \right )\},
\end{eqnarray}

and
\begin{eqnarray}
n_i(\Delta E)= \left( B- ( m^{(i)}_{\beta})^2 \right)^{3/2} \Theta \left( \Delta E-(m_{\beta}^{(i)}-m_1) \right),
\end{eqnarray}

with
\begin{eqnarray}
B=\Delta E ( \Delta E+2m_1). \nonumber
\end{eqnarray}

\begin{figure}[h]
\begin{center}
\epsfig{figure=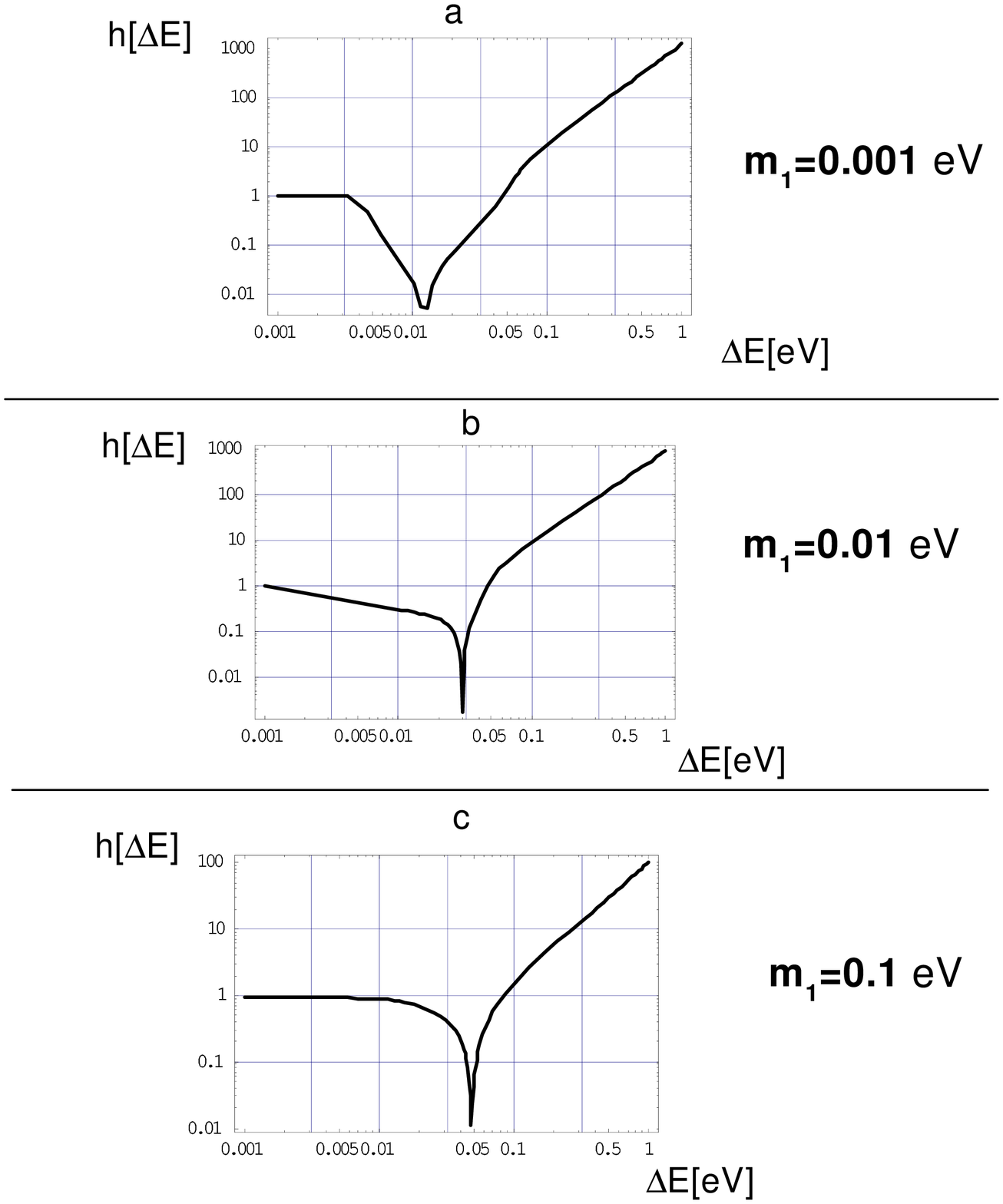, scale=0.7}
\caption{h($ \Delta E$) as a function of $\Delta  E$ for minimal neutrino mass (a) $m_1=0.001 \ eV$ , (b) $m_1=0.01 \ eV$ , (c) $m_1=0.1 \ eV$ .}
\label{spect}
\end{center}
\end{figure}

To compare both approximate spectra the ratio 
\begin{equation}
h( \Delta E)=\frac{|n_0(\Delta E)-n_2(\Delta E)|}{|n_0(\Delta E)-n_1(\Delta E)|}
\end{equation}
is plotted on Fig.~\ref{spect}
for three different neutrino masses $m_1=0.001 \ eV, m_1=0.01 \ eV $ and $m_1=0.1 \ eV$. 
We can see that  independently of chosen $m_1$ and for $ \Delta E>m_3-m_1$,  $h( \Delta E) > 1$. 
We know that $m_{3}-m_{1} <0.08 \ eV$ \cite{ost}. It will be very difficult to get such a small  energy spectrum resolution. 
So, let us conclude. In practice $ \Delta E \gg m_3-m_1$, and approximate spectrum with $m_{\beta}=\sqrt{\sum|U_{ei}|^2 m_i^2}$ should be used 
in  future searches of neutrino masses in the tritium $\beta$ decay.

\newpage

\newpage

{\bf Acknowledgments}
This work was supported by the Polish Committee for Scientific Research under Grant Nos. 2P03B05418 and 5P03B08921.


\begin{thebibliography}{99}
\bibitem{neutphys} see e.g W. Kuandig et al., in ``Neutrino Physics'' ed. by K. Winter, Cambridge Univ. Press 1991, p. 144; F. Boehm, P. Vogel `` Physics of massive neutrinos'', Published by Press Syndicate of the University of Cambridge 1987, 1992 .
\bibitem{mainz}J. Bonn, et al, Nucl. Phys.B {\bf 91} (2001)273.
\bibitem{troisk}V. M. Lobashev,  Nucl. Phys.B {\bf  91} (2001)273.
\bibitem{eg}R. E. Shrock, Phys. Lett.B {\bf 96} (1980) 159.
\bibitem{katrin}V. Aseev et al. http://www.hep.anl.gov/ndk/hypertext/mumi/html, A. Osipowicz at.al. [hep-ex/0109033]
\bibitem{vissani}F.~Vissani,
hep-ph/0102235.
\bibitem{farzan}Y.~Farzan, O.~L.~Peres and A.~Y.~Smirnov,
hep-ph/0105105.
\bibitem{gonzales}
M.~C.~Gonzalez-Garcia, M.~Maltoni, C.~Pena-Garay and J.~W.~Valle,
Phys.\ Rev.\ D {\bf 63}, 033005 (2001)
[hep-ph/0009350].
\bibitem{toshito}T.~Toshito  [SuperKamiokande Collaboration],
hep-ex/0105023.
\bibitem{ost}M. Czakon, J. Gluza, J. Studnik, M. Zra\l ek Acta. Phys. Pol. B {\bf 31}, 1365 (2000)
\end{thebibliography}
\end{document}